\documentclass[11pt]{article}

\pdfoutput=1
\usepackage[pdftex]{color}
\usepackage{amssymb}
\usepackage{amsthm}
\usepackage{amsmath}
\usepackage{latexsym}
\usepackage{amscd}
\usepackage{graphicx}
\usepackage[pdftex, colorlinks=true, citecolor=green]{hyperref}
\usepackage{lscape}
\usepackage{setspace}
\usepackage{multirow}

\newcommand{\ds}{\displaystyle}

\newcommand{\ben}{\begin{equation}}     
\newcommand{\eeqn}{\end{equation}}
\newcommand{\bey}{\begin{eqnarray}}
\newcommand{\eey}{\end{eqnarray}}

\begin{document}

\noindent {\Large
\textbf{Nonlinear network dynamics under perturbations of the\\  underlying graph}
}
\\
\vspace{4mm}
 Anca R\v{a}dulescu$^{*,}\footnote{\doublespacing Assistant Professor, Department of Mathematics, State University of New York at New Paltz; New York, USA; Phone: (845) 257-3532; Email: radulesa@newpaltz.edu}$, Sergio Verduzco-Flores$^{2}$
\\
\indent $^1$ Department of Mathematics, SUNY New Paltz, NY 12561
\\
\indent $^2$ Department of Psychology,  University of Colorado at Boulder, CO 80309
\\

\begin{abstract}
\noindent Many natural systems are organized as networks, in which the nodes (be they cells, individuals or populations) interact in a time-dependent fashion. The dynamic behavior of these networks depends on how these nodes are connected, which can be understood in terms of an adjacency matrix, and connection strengths. The object of our study is to relate connectivity to temporal behavior in networks of coupled nonlinear oscillators. We investigate the relationship between classes of system architectures and classes of their possible dynamics, when the nodes are coupled according to a connectivity scheme that obeys certain constrains, but also incorporates random aspects.\\

\noindent We illustrate how the phase space dynamics and bifurcations of the system change when perturbing the underlying adjacency graph. We differentiate between the effects on dynamics of the following operations that  directly modulate network connectivity: (1) increasing/decreasing edge weights, (2) increasing/decreasing edge density, (3) altering edge configuration by adding, deleting or moving edges.\\

\noindent We discuss the significance of our results in the context of real life networks. Some interpretations lead us to draw conclusions that may apply to brain networks, synaptic restructuring and neural dynamics.\\
\end{abstract}

\clearpage
\noindent {\bf The study of a dynamical system with interconnected nodes tends to gain little insight from the graph-theoretical properties of the underlying graph. Brain networks show properties such as small-world connectivity and repeated motifs, but the computational impact of those properties remains unclear. An avenue pursued in this paper in order to investigate the relation between a network's structure and its dynamics is to find relations between the network's underlying graph, and the behavior of this system. We perform a computational study of dynamics under different forms of connectivity between two densely connected modules. Using phase diagrams and a probabilistic extension of bifurcation diagrams in the parameter space, we find several properties of the network dynamics. Among them, we find that the spectrum of the adjacency matrix is a poor predictor of dynamics when using nonlinear nodes, that increasing the number of connections between the two nodes is not equivalent to strengthening a few connections, and that there is no single factor among those we tested that governs the stability of the system.} \\

\section{Introduction}

A large body of literature over the past decade has been dedicated to the study
of networks and their applications to understanding the behavior of social,
neural and biological systems. One of the particular points of interest has been
the question of how the \emph{hardwired structure} of the network (its
underlying graph) affects its \emph{function}, for example in the context of
information storage or transmission between nodes along
time~\cite{boccaletti2006complex}. There are two key coupling aspects that
govern dynamic function in such networks: the underlying graph (characterized by
its adjacency matrix) and the connection strengths. Understanding the effects of
configuration (which is another term we'll use for the adjacency matrix) on
coupled dynamics is of great importance for a wide variety of applications. 

There are not many previous studies dealing with the direct effect of configuration on
a dynamical system. Naquib et al~\cite{naqib2012tunable} found that by varying
the location of synthetic sites for two different chemical species (one
excitatory and one inhibitory) that diffuse across a one-dimensional ring, they
could find many dynamic behaviors, including fixed points, out-of-phase
oscillations, quasiperiodicity, and chaos. An interesting aspect of this study
is that the structure of the system (i.e. the location and identity of the
synthetic sites) acts as a bifurcation parameter. We also explore the idea of
having structure as a bifurcation parameter; a structure found in the adjacency
matrix of the network.

The differential equations that model dynamical systems consisting of
interconnected nonlinear nodes do not usually admit closed form analytical
solutions. In this situation, the qualitative behavior of the system may 
still be grasped through bifurcation analysis, which for complex systems is
usually carried out using numerical continuation methods. These methods are
appropriate to study a single system, where the graph describing the 
connections among its nodes is fixed.  For the aim of this paper, however, 
we want to understand how the system's dynamics change as the adjacency
matrix that describes its underlying graph experiences variations. 
This involves evaluating a large number of systems, each with a different 
adjacency matrix. Because of this, bifurcation analysis with numerical
continuation methods becomes computationally expensive (the number of
possible adjacency matrices increases exponentially with the number
of nodes), and it would be unclear how to interpret a large number of 
bifurcation diagrams, each one for a different underlying graph.

We propose a simple approach to visualize the qualitative behavior of
nonlinear dynamical sytems with an underlying graph structure. The
approach starts by discretizing the values in the bifurcation diagram
to obtain a finite number of points in parameter space. For each point
in parameter space we take a sample of the adjacency matrices, and
for each one, find whether the system expresses the dynamical behaviors
of interest (e.g. bistability, or oscillations). For each dynamical behavior 
of interest, we can create
a diagram where each point in parameter space is associated with the
fraction of adjacency matrices causing the behavior to appear in the
system. This diagram expresses, for each point in parameter space, what
is the probability that a given dynamical behavior will appear in the
absence of information about the system's configuration. If the sampling of adjacency
matrices is restricted to those satisfying a particular constraint
(e.g. a fixed number of ones in the adjacency matrix), then the diagram
will express an approximation to the corresponding conditional probabilities.
We use the term of \emph{(dynamic) behavior frequency plots} to refer to diagrams produced this way, which can be applied for any specific behavior.

To focus in a particular direction, we place and interpret our results in the context of brain architecture and dynamics. Understanding the way in which various parts of the brain (from the micro-scale of neurons to the macro-scale of functional regions) are wired together is one of the great scientific challenges of the 21st century, currently being addressed by large-scale research collaborations, such as the Human Connectome Project~\cite{toga2012mapping}. Many recent studies (e.g., ~\cite{sporns2011human,sporns2001classes,park2013structural}) have used a combination of dynamical systems and graph theoretical approaches to investigate general organizational principles of brain networks. With nodes and edges defined according to imaging modality appropriate scales, empirical studies have found certain generic topological properties of the human brain architecture, such as modularity, small-worldness, the presence of rich clubs and hubs, and other connectivity patterns~\cite{bullmore2009complex,rubinov2010complex,park2013structural,sporns2004motifs}. 

Purely empirically-based analyses cannot, however, explain in and off themselves the mechanisms by which connectivity patterns may actually act to change the system's dynamics, and thus the observed behavior. Substantial research efforts are being directed towards constructing underlying network \emph{models} that are tractable theoretically or numerically, and which could therefore be used in conjunction with the data towards interpreting the empirical results, and for making further predictions. To this aim, the theoretical dependence of dynamics on connectivity (e.g., in the context of stability and synchronization in networks of coupled neural populations) has been investigated both analytically and numerically, in a variety of contexts, from biophysical models~\cite{gray2009stability} to simplified systems~\cite{siri2007effects}. These analyses revealed a rich range of potential dynamic regimes and transitions~\cite{brunel2000dynamics}, shown to depend as much on the coupling parameters of the network as on the arrangement of the excitatory and inhibitory connections~\cite{gray2009stability}. Understanding and teasing apart the different effects of these dependences is the central goal of this work.

\begin{figure}[h!]
\begin{center}
\includegraphics[scale=0.35]{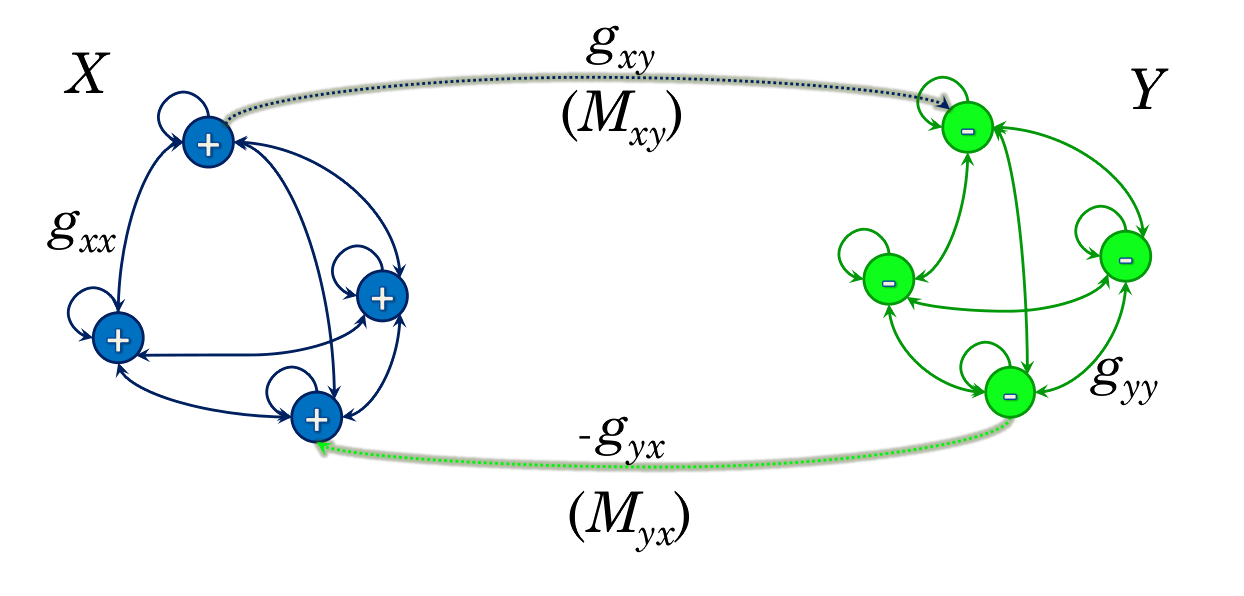}
\end{center}
\caption{\doublespacing \emph{\small {\bf Schematic representation of the network for $N=4$ nodes per module.} Module $X$ is shown on the left; module $Y$ is shown on the right; they are both fully-connected, local sub-graphs of the oriented graph corresponding to the whole network. The thick blue arrow shows that there are $M_{xy}$ connections from $X$ to $Y$, and the thick green arrow that there are $M_{yx}$ connections from $Y$ to $X$ (the edges are not shown in their specific positions, making this a general representation of any configuration with the given edge densities). The coupling weights $g_{xx}$, $g_{xy}$, $g_{yx}$, $g_{yy}$ are marked on the corresponding edges.}}
\label{network}
\end{figure}

We will start our study by considering a type of architecture already used in previous work~\cite{radulescu2013network,radulescu2014network}: an oriented graph composed of two interconnected cliques (fully connected subgraphs), module $X$ and module $Y$, so that all nodes $\{x_k\}_{k=\overline{1,N}}$ within $X$ are mutually connected by ``excitatory'' edges with equal positive weights $g_{xx}$, and all nodes $\{y_k\}_{k=\overline{1,N}}$ within $Y$ are mutually connected by excitatory edges with positive weights $g_{yy}$ (Figure~\ref{network}). The connectivity patterns from $X$-to-$Y$ and $Y$-to-$X$ can be described by two binary $N \times N$ blocks $A=\left( A_{kp} \right)$ and $B=\left( B_{kp} \right)$, representing which of the nodes in $X$ are cross-connected to nodes in $Y$ (with equal excitatory, positive weights $g_{xy}$) and conversely, which of the nodes in $Y$ are connected to nodes in $X$ (with inhibitory, negative weights $-g_{yx}$). This graph structure was chosen in previous work as a very simple framework for studying the excitatory/inhibitory feedback interaction in a control system composed of two brain regions (in our case the amygdala and the prefrontal cortex), with the nodes representing hemodynamic oscillators. The set-up can be used, however, at other spacial and temporal scales, or for any bimodular network defining a similar feedback loop. One can easily adapt it to incorporate more than two modules, or can prune out the dense intra-modular connections to obtain more realistic conditions, while keeping it simple enough to address numerically or analytically, for sufficiently large numbers of nodes.

In a previous paper~\cite{radulescu2014network}, we had focused primarily on the properties of the graph underlying a neural network, and discussed how factors such as changes in density or other edge restructuring may affect the spectrum of the adjacency matrix. In this paper, our attention is directed towards further relating adjacency properties to the system's temporal behavior, and understanding the subsequent changes they trigger in the coupled dynamics. More precisely, we are interested in varying the number of active inter-modular edges $M_{xy}$ and $M_{yx}$ (i.e., $M_{xy}$ is the number of $1$ entries in $A$ and $M_{yx}$ is the number of $1$ entries in $B$, both ranging from zero to the theoretical maximum $N^2$), but also in changing the edge configuration for a fixed pair $(M_{xy}, M_{yx})$ (which we will call the \emph{density type} of the graph, for the remainder of this paper). 

We investigate the consequences that each of these two aspects has on the overall dynamics of a system of coupled nodes,  where we identify each node with a continuous-time nonlinear oscillator. From a vast collection of such models, we drew our inspiration from the simple and traditional  Wilson-Cowan equations, a system conceived and used historically to model interaction of excitatory and inhibitory neural populations~\cite{wilson1972excitatory}. This two-dimensional system was shown to exhibit interesting dynamic behavior in the two-dimensional phase-space, with Hopf and fold bifurcations between stable equilibria and stable limit cycles (including bistability windows). In our study, we work within the parameter ranges proposed in the original Wilson-Cowan paper~\cite{wilson1972excitatory} as weel as in subsequent work in higher dimensions~\cite{borisyuk1995dynamics,campbell1996synchronization}, thus placing the system in the vicinity of the interesting phase-plane phenomena. We study how the phase-plane dynamics and the parameter-plane transitions change when perturbing the underlying coupling graph.

Some of the results we obtained for this system were intuitive, but others were rather unexpected. For example, we established  that the eigenvalues of the adjacency matrix do not determine the dynamic behavior, which is not surprising. Conversely however, the dynamic behavior seems to determine the eigenvalues. The intermodular connection weights can put the system in sensitive regimes where changes in the adjacency matrix is more likely to affect the dynamic behavior. (At least in the low dimensional case, these regimes appear for the $g_{xy},g_{yx}$ values near bifurcation curves for individual configurations.) In fact sparse connectivity (smaller $M_{xy}, M_{yx}$ values) promotes a higher variety of dynamic behaviors, accessible by changes in either weights of adjacency configuration. Quite surprisingly, we found that the network does not experience chaotic dynamics.

The simplicity of the system makes it ideal for analytical and numerical investigations. However, its tight intramodular coupling (leading to a high degree of synchronizations in the nodes' activity) and its lack of aperiodic behavior make it unrealistic as a model of real worls networks, which are typically more complex and may spend considerable time in chaotic regimes. To address this, we considered in Section~\ref{coupled_WC_section} an extended model of coupled Wilson-Cowan pairs. While this system also illustrates the tight relationshops between connectivity and dynamics, its behavior is much richer; one can easily produce desynchonization and/or tune the system to aperiodic behavior by changes in its parameters.


\section{Coupled nonlinear oscillators}
\label{nonlinear_osc}

We consider the following 2N-dimensional system of nonlinear oscillators (whose architecture is illustrated in Figure~\ref{network} when $N=4$):

\begin{eqnarray}
\dot{x}_k &=& \ds -x_k + (1-x_k) \cdot {\cal S}_{b_x,\theta_x} \left(-\sum_{p=1}^{N}{g_{yx} a_{kp} y_p} + \sum_{p=1}^{N}{g_{xx} x_p} + P \right) \nonumber \\
\dot{y}_k &=& \ds -y_k + (1-y_k) \cdot {\cal S}_{b_y,\theta_y} \left( \sum_{p=1}^{N}{g_{xy} b_{kp} x_p} + \sum_{p=1}^{N}{g_{yy} y_p} + Q \right)
\label{mothersys}
\end{eqnarray}

\noindent with $1 \leq k \leq N$. Each node is driven by external sources ($P$ for the nodes $x_k$ in the module $X$, and $Q$ for the nodes $y_k$ in the module $Y$). In addition, each node receives input from all other nodes that are connected to it through incoming edges, with weights $g$, indexed as described in the previous section and in Figure~\ref{network}. The coefficients $a_{kp},b_{kp} \in \{ 0,1 \}$ are the binary entries of the adjacency blocks $A$ and $B$. The effective input to each node is the sum of all such external and internal sources, modulated by the sigmoidal:

\begin{equation}
{\cal S}_{b,\theta}[Z] = \frac{1}{1+\exp(-b[Z-\theta])}-\frac{1}{1+\exp(b\theta)}
\end{equation}

\noindent with parameters $b=b_x$ and $\theta = \theta_x$ when the target node is in module $X$, and $b=b_y$ and $\theta = \theta_y$ when the target node is in module $Y$. Throughout our analysis, we fixed: $b_x=1.3$, $b_y=2$, $\theta_x=4$, $\theta_y=3.7$, $g_{xx}=16/N$, $g_{yy}=3/N$, $P=1.5$, $Q=0$. We allowed the range $[0,30]$ for the $X$-to-$Y$ and $Y$-to-$X$ connectivity strengths $g_{xy}$ and $g_{yx}$. The form of the equations and the parameters are typical for Wilson-Cowan dynamics. 

A comprehensive study of parameter dependence for such a system would be almost intractable (as it would be for any system attempting to model real world, complex phenomena affected by a wide collection of factors). Let us notice, for example, that perturbing the individual node dynamics (e.g., the logistic function) has its own -- distinct -- effect on the temporal evolution of the coupled system. The best one can do is to analyze the sensitivity of the system with respect to one or two parameters of interest at the time, and eventually use this information to quantify and directly compare the effects of each factor on the system's behavior.\\

\noindent To continue, we will first consider a small network size ($N=2$), and inspect the dynamic behavior of the system for every possible theoretical configuration (adjacency matrix) corresponding to a fixed pair of edge densities. In Section~\ref{2Dsection}, we discuss the cases $(M_{xy},M_{yx})=(3,3)$ and $(M_{xy},M_{yx})=(2,3)$, but a similar analysis can be carried for any density pair. We study how small changes in the graph (such as adding/deleting an edge, or moving an edge by a sequence of add/delete operations) influence the system's dynamics, and we try to understand in which scenario these dynamics are most sensitive to weight changes. Furthermore, we are interested in finding whether structural changes (edge shifting) may have comparable effects with varying the weights, or under which circumstances this may be true. 

Our specific interest remains, however, in studying what happens for higher network sizes $N$. That is because natural systems are likely to be formed, even at a macroscopic level, of hundreds or thousands of node-units. Since the number of configurations increases extremely fast (combinatorially squared) with the size $N$, it is no longer ideal, for high $N$ values, to describe each individual configuration in this large set; we propose a probabilistic approach to be more appropriate. In Section~\ref{hiDsection}, we define \emph{behavior frequency plots}, quantifying the statistical likelihood of the system (over the distribution of all possible configurations corresponding to a fixed density type) to exhibit a certain dynamics at a fixed combination of edge weights. While in this paper we only establishe a proof of principle, by inverstigating small sizes ($N=4$, leading to thousands of configurations for each density type), the methods can be applied to higher network sizes by using increased resources, or by concentrating the search on more specific aspects.\\ 

\noindent Through the following sections, we will use the notation ${\cal D}^{M_{xy},M_{yx}}$ for the collection of all adjacency matrices with density type $(M_{xy},M_{yx})$.


\subsection{Low dimensional dynamics. Bifurcation diagrams}
\label{2Dsection}

In the low dimensional case of $N=2$ nodes per module, the system \eqref{mothersys} becomes:
\begin{eqnarray}
\dot{x}_1 &=& \ds -x_1+(1-x_1) \cdot {\cal S}_{b_x,\tau_x}[g_{xx}(x_1+x_2)-g_{yx}(a_{11}y_1+a_{12}y_2)+P] \nonumber \\
\dot{x}_2 &=& \ds -x_2+(1-x_2) \cdot {\cal S}_{b_x,\tau_x}[g_{xx}(x_1+x_2)-g_{yx}(a_{21}y_1+a_{22}y_2)+P] \nonumber \\
\dot{y}_1 &=& \ds -y_1+(1-y_1) \cdot {\cal S}_{b_y,\tau_y}[g_{xy}(b_{11}x_1+b_{12}x_2)-g_{yy}(y_1+y_2)+Q] \nonumber\\
\dot{y}_2 &=& \ds -y_2+(1-y_2) \cdot {\cal S}_{b_y,\tau_y}[g_{xy}(b_{21}x_1+b_{22}x_2)-g_{yy}(y_1+y_2)+Q]
\label{2Dsystem}
\end{eqnarray}

\noindent Intuitively, we expect the dynamics to be influenced by the flow/dissipation of the information in the system, i.e., by the average length of the minimal path that connects any two nodes. While the density type $(M_{xy},M_{yx})$ strongly influences the dynamics of the system, it clearly does not completely determine temporal behavior in and off itself, and the dynamics are only partly encoded in the density type, or in the adjacency spectrum. One common sense expectation is that, for a fixed density type $(M_{xy},M_{yx})$, two adjacency configurations with the same eigenspectrum can produce significantly different phase-space dynamics. We verify this conjecture and try to better describe the correspondence between adjacency and dynamics, but we also propose that other options for measuring the properties of the graph may capture better the system's dynamic complexity.

For a phase-plane analysis of a nonlinear dynamical system, one typically starts by establishing the position and stability of  equilibria, searching for invariant sets (e.g. cycles, invariant tori, etc) and for potential aperiodic/chaotic behavior. Since, due to the nonlinearity of the system, describing these objects precisely is quite challenging, we use numerical algorithms to approximate the attractors' position and shape, establish their stability and study their change under perturbation of parameters. Throughout this study, we keep all other system parameters fixed, and only vary the between-module connection strengths $g_{xy}$ and $g_{yx}$, and the system's underlying geometry (by allowing various configurations for the binary matrices $\ds A= \left[ \begin{array}{cc} a_{11} & a_{12}\\ a_{21} & a_{22} \end{array} \right]$ and $\ds B= \left[ \begin{array}{cc} b_{11} & b_{12}\\ b_{21} & b_{22} \end{array} \right]$). This choice is motivated by our aim to understand and compare the different effects on dynamics of three distinct ways of altering inter-modular connectivit: (1) by changing the edge density type $(M_{xy},M_{yx})$, (2) by changing the node-node edge configuration (the positions of the 1 entries in the binary matrices $A$ and $B$) and (3) by changing the inter-modular edge weights ($g_{xy}$ and $g_{yx}$).

In order to understand, for each individual adjacency configuration, the changes in dynamics produced by varying $g_{xy}$ and $g_{yx}$, we first use bifurcation diagrams in the $(g_{xy},g_{yx})$ parameter plane. Then, we observe how these diagrams change when perturbing the underlying adjacency graph. To generate the bifurcation diagrams, we used the continuation algorithms provided by the Matcont package~\cite{Matcont}, initialized in a region containing values of $g_{xy}$ and $g_{yx}$ corresponding to Hopf and saddle node bifurcations in the classic Wilson-Cowan system. We investigated the Hopf and limit point (saddle point) curves in our own coupled system, delimiting behaviors such as convergence to a unique stable equilibrium versus oscillations towards a stable limit cycle (including bistability).

To illustrate our ideas, the two tables in the Appendix show all possible $(g_{xy},g_{yx})$ parameter planes that can be obtained for $N=2$ and density types $(M_{xy},M_{yx})=(3,3)$ and $(M_{xy},M_{yx})=(2,3)$, respectively.  Interestingly, all 16 combinatorial configurations in ${\cal D}^{3,3}$  produce only four distinct dynamic parameter planes, which we will call the dynamic classes for $(M_{xy},M_{yx})=(3,3)$,a dn which we show in Table~\ref{table_3_3}. Notice that all 4 dynamic classes can be obtained by fixing $A$ to any configuration and considering all 4 cases for $B$s, but also by fixing $B$ and considering all 4 configurations for $A$. Similarly, all 24 combinatorial configurations in ${\cal D}^{2,3}$ produce only six dynamic classes, distinct than the ones in ${\cal D}^{3,3}$, shown in Table~\ref{table_2_3}, all obtainable by fixing $B$ and varying $A$.

The presence of bifurcations for all dynamic classes implies that, when fixing the network, changing one of the weights $g_{xy}$ or $g_{yx}$ can push the system over a bifurcation curve, placing it in a different regime. This change may consist for example of switching between ``rest'' (convergence to a stable equilibrium) and ``oscillations'' (convergence to a limit cycle) when crossing a Hopf bifurcation, or of sharply switching attractors (when crossing a limit point curve). 

Aternatively, looking across all dynamic classes for each $(M_{xy},M_{yx})$, one may easily note that the changes in dynamics triggered by changes in configuration are rather localized to certain regions in the parameter plane. That is, the behavior of the system might be, between classes, very different or very similar at different points in the $(g_{xy},g_{yx})$ plane. This suggests that the system's sensitivity to the network geometry depends on the actual connection weights. There appears to be a critical $(g_{xy},g_{yx})$ locus were the system is most sensitive to geometry: deleting or shifting one edge can push the system from a stable equilibrium (in one panel) to oscillations (in a different panel). Away from this region, there is a more topographic correspondence between parameter planes (i.e., the dynamic classes have qualitatively more consistent, or even identical behavior between panels).\\

\noindent We say that two configurations are in the same adjacency class if they have the same eigenspectra. When investigating the relationship between the adjacency configuration and the dynamic behavior of the system, a natural question to ask is whether dynamic classes may be predicted simply by looking at the adjacency spectrum.  We conjecture that the correspondence dynamics $\rightarrow$ adjacency classes is well defined, but clearly not bijective. That is: a specific  dynamic class can't be obtained from two different configurations, but a single adjacency class may lead to different dynamics.

While in general, for high dimensions, proving this relationship may be quite difficult, for low dimensions it is easy to illustrate. For example, Table~\ref{table_3_3} shows each configuration in ${\cal D}^{3,3}$ together with its adjacency and dynamic class. In this case, there are three distinct adjacency eigenspectra (designated by letters ${\cal A}$ through ${\cal C}$),  each class containing respectively 8, 4 and 4 of the total of $(2N)^2 = 16$ configurations. In counterpart, there are four distinct dynamics classes (designated by indices $i$ through $iv$). With this convention, the table shows that no dynamics can be obtained from multiple adjacency classes, but that some adjacency classes can lead to multiple dynamics. Similarly, Table~\ref{table_2_3} shows how the 6 dynamic classes are mapped to the 4 adjacency classes in the case of ${\cal D}^{2,3}$.

This suggests that, while the adjacency class, together with the density type, clearly have a contribution to dynamics, they cannot be directly used either to predict these dynamics. In our current work, we are investigating whether other descriptions of the adjacency matrix are better choices to help predict the dynamics or the complexity of a network's evolution. Node degree distribution, connectivity coefficient, number of particular motifs may be finer network measures than edge density, or adjacency spectrum, and therefore more efficient in classifying dynamic complexity.\\

\begin{figure}[h!]
\begin{center}
\includegraphics[width=.8\textwidth]{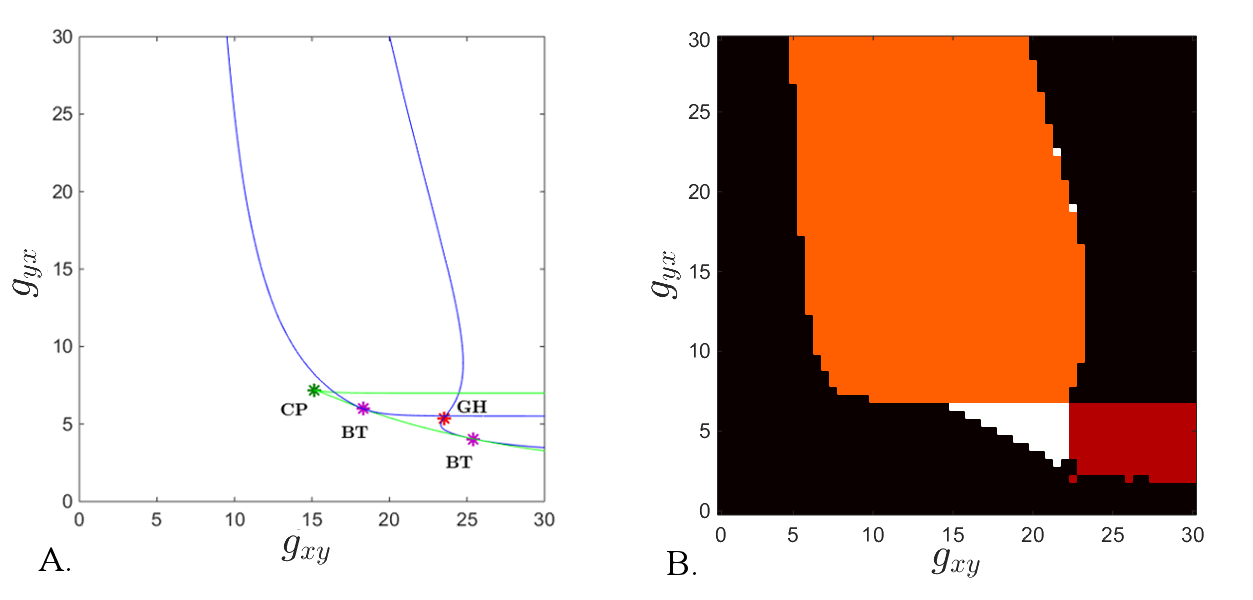}
\caption{\doublespacing \label{single_N2} \emph{\small  {\bf Bifurcation diagrams versus search algorithm.} {\bf A.} Bifurcation diagram in the  $(g_{xy},g_{yx})$ plane for the dynamic class $(ii)$ in ${\cal D}^{2,3}$, created with the Matcont extension algorithms. Hopf curves are shown in blue, limit point curves in green, and codimension two points are shown as stars: green (cusp points), red (generalized Hopf points), purple (Bogdanov-Takens points). {\bf B.} Dynamic regimes in the  $(g_{xy},g_{yx})$ plane for the same dynamic class, obtained using our numerical search for different behaviors within the system: the locus corresponding to a unique stable equilibrium is in black, the locus for multiple stable equilibria is in red, the locus for a unique stable cycle is in orange, and the locus where the stable equilibrium and the stable cycle coexist is in white.}}
\end{center}
\end{figure}

\noindent It is becoming clear that, even for small $N$, the system has many dynamic possibilities (depending on configuration), thus making undesirable an individual descriptive approach to each configuration-specific parameter space. A statistical approach seems more appropriate, bearing in mind that some dynamic classes may be more substantial than others, and thus have a stronger contribution to driving these statistics. While these are ideas that we elaborate more in the following sections, here we set the grounds for this path by describing the numerical methods used and by illustrating how they work on a simple $N=2$ example.\\

\noindent For the each $(g_{xy},g_{yx})$ parameter point we took a sample of adjacency matrices with a given density. For each adjacency matrix in the sample we ran simulations and analyzed each one in order to find the range of dynamic behaviors it could produce when starting from different initial conditions. Our search algorithms could detect six types of behaviors: 1) a single fixed point, 2) multiple fixed points, 3) periodic oscillations, 4) non periodic oscillations, 5) both a single fixed point and periodic oscillations, and 6) both multiple fixed points and periodic oscillations. We only analyzed the second half of the simulations in order to remove the transient part of the activity. 

For each pair of connection weights and for each adjacency matrix we explored the space of initial conditions using a basic Particle Swarm Optimization (PSO) algorithm. The utility function used by the PSO algorithm depend on which behaviors had already been found. If only a single fixed point had been found then the initial condition with the largest utility was the one with largest amplitude in its oscillations. This amplitude was determined as the difference between the mean value of each node's response and its largest value in the last quarter of the simulation. The largest amplitude among all nodes was selected. If no fixed points had been found, then the utility function was set to one minus the utility of the previous case.

Detecting fixed points was done using the same amplitude that constituted the utility function for the PSO algorithm. When this amplitude was below a threshold, a fixed point was detected. To detect multiple fixed points, for each initial condition where a fixed point was found the average value of the response for the first node was stored. If the difference between the largest stored average value and the smallest one was above a threshold, multiple fixed points were detected.

Detecting periodic oscillations was done using a basic algorithm that convolved the time-discretized response of a node with with a time inverted version of itself. Intuitively, this response could be conceived as a vector, and the convolution as an inner product between a part of this vector and a shifted version of itself. The reason why this algorithm works is that when the sections of the vectors participating in the inner product are normalized, the inner product will attain its maximum value when the response vector and its time shifted version are the same. This happens when the response signal is periodic.

A non periodic oscillation was detected when the response was not a fixed point, but our algorithm could not detect periodic behavior. Whenever aperiodic behavior was detected, the simulation was extended for a longer period of time and then analyzed again in order to prevent false detections due to transient properties of the response.\\

\noindent We first illustrate the efficiency of this search algorithm by applying it to find all behaviors in the parameter plane for the configuration $\ds A= \left[ \begin{array}{cc} 1 & 0\\ 1 & 0 \end{array} \right]$ and $\ds B= \left[ \begin{array}{cc} 0 & 1\\ 1 & 1 \end{array} \right]$ in ${\cal D}^{2,3}$, which was found by Matcont to be of class $(ii)$. Figure~\ref{single_N2} compares side by side the diagrams obtained in this particular case: via the Matcont software on the left, and via our search algorithm on the right.

\begin{figure}[h!]
\begin{center}
\includegraphics[width=\textwidth]{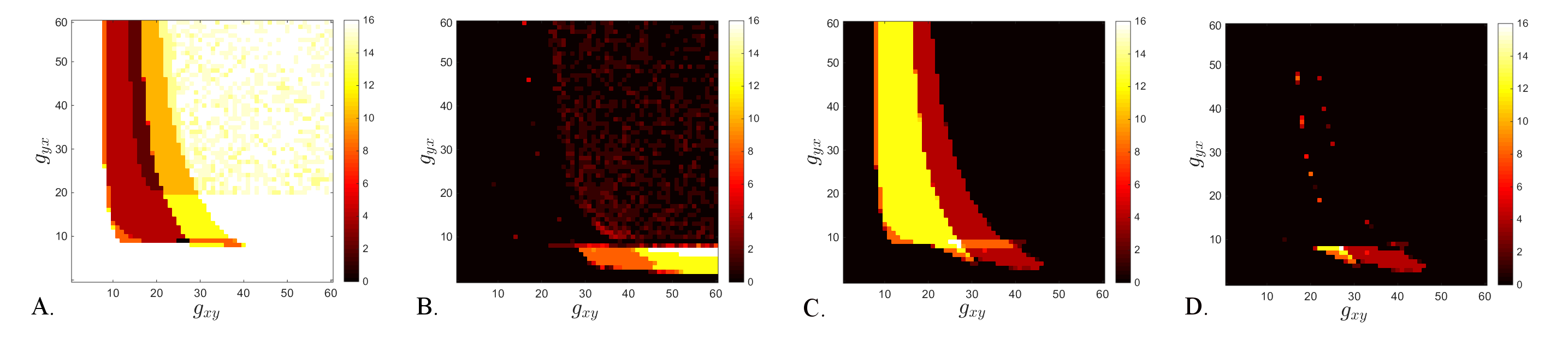}
\caption{\doublespacing \label{bifurcations_N2} \emph{\small  {\bf  Behavior frequency plots}, showing the number (out of all 16 configurations in ${\cal D}^{3,3}$) which exhibit: {\bf A.} one stable equilibrium; {\bf B.} multiple stable equilibria;  {\bf C.} one stable cycle; {\bf D.} a coexisting stable cycle and equilibrium.}}
\end{center}
\end{figure}

We then used the search algorithm by itself, to illustrate the likelihood for each behavior at each parameter point $(g_{xy},g_{yx})$, over all configurations in ${\cal D}^{3,3}$. We will call the parameter loci for different behaviors -- p-bifurcations of the system. Each panel in Figure~\ref{bifurcations_N2} illustrates the likelihood for an arbitrary configuration in ${\cal D}^{3,3}$ to exhibit one of the following attracting sets: a globally stable equilibrium (Figure~\ref{bifurcations_N2}a), multiple stable equilibria (Figure~\ref{bifurcations_N2}b), a globally stable limit cycle (Figure~\ref{bifurcations_N2}c), or a coexisting stable equilibrium and stable cycle (Figure~\ref{bifurcations_N2}d). Our search algorithm found only artifactual aperiodic behavior, which, upon inspection, was clearly due to a slower initial transient phase of the solution, mistakenly labeled by our code as aperiodic behavior.

In higher dimensions, one may expect the system's attractors to transcend simple limit cycles  (for example, a paper by Borisyuk et al~\cite{borisyuk1995dynamics} found a similar four-dimensional, coupled system to additionally exhibit symmetric, antisymmetric and nonsymmetric invariant tori), which are hard (and computationally rather expensive) to track down. One of our goals is to investigate the presence of aperiodic behavior in our system for higher dimensions, which we do in the next section for the case $N=4$.

\subsection{P-bifurcations}
\label{hiDsection}

In this section, we focus on constructing and understanding behavior frequency plots. Each point $(g_{xy},g_{yx})$ in the parameter plane may correspond or not, for each adjacency configuration in ${\cal D}^{M_{xy},M_{yx}}$ , to a specific dynamic behavior. In other words the point will be on one side versus the other of some bifurcation curve in $(g_{xy},g_{yx})$, with a specific probability (over the whole configuration distribution ${\cal D}^{M_{xy},M_{yx}}$ ). This represents in a sense a probabilistic extension of the concept of bifurcation,

As shown before, one can define the p-bifurcation diagram of a system for any particular dynamic behavior. Fix the size $N$ and the density type $(M_{xy},M_{yx})$. For each pair of edge weights $(g_{xy},g_{yx})$, we can calculate (or estimate numerically) the fraction ${\cal P}(g_{xy},g_{yx})$ of adjacencies in ${\cal D}^{M_{xy},M_{yx}}$ which, for weights $(g_{xy},g_{yx})$, lead to a specific dynamic behavior. E.g., by estimating the fraction of configurations which lead to coexistence of a stable equilibrium and a stable cycle, one can establish the locus in the parameter plane where there exist configurations with equilibrium/cycle bistability, evaluate how likely it is to randomly pick a configuration with such bistable behavior,  and observe what is needed to push the system from a regime of likely bistability into a purely oscillatory or quiet regime.

For $M_{xy}=M_{yx}=N^2$, there is only one possible configuration, and the behavior loci are delimited by regular bifurcation curves. When $\lvert {\cal D}^{M_{xy},M_{yx}} \rvert \neq 1$, the transition is smooth, so that there is a region where $0 < {\cal P} <1$, which corresponds to a ``smeared'' bifurcation curve.

\begin{figure}[h!]
\begin{center}
\includegraphics[width=\textwidth]{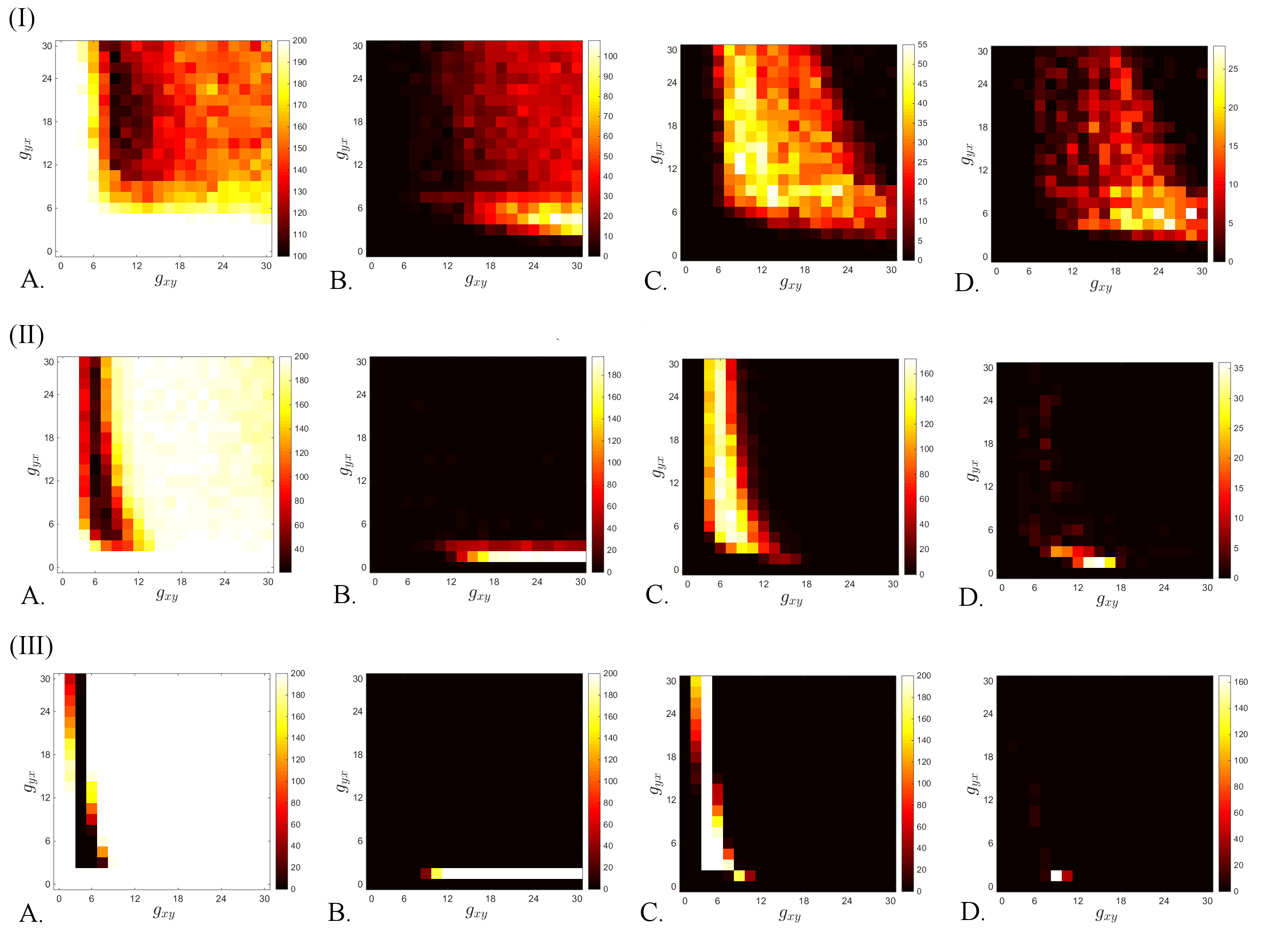}
\caption{\doublespacing \label{bifurcations_N4_identical} \emph{\small  {\bf  Behavior frequency plots for $N=4$, for equal densities $M_{xy}$ and $M_{yx}$.} We show the fraction of all configurations in {\bf (i)} ${\cal D}^{4,4}$; {\bf (ii)} ${\cal D}^{8,8}$; {\bf (iii)} ${\cal D}^{12,12}$ which exhibit: {\bf A.} one stable equilibrium; {\bf B.} one stable cycle; {\bf C.} multiple stable equilibria; {\bf C.} a coexisting stable cycle and equilibrium. No other behaviors were found. The simulations are based on a randomly generated sample of $S=200$ configurations in the respective ${\cal D}^{M_{xy},M_{yx}}$ (hence the color bar represents numbers from 0 to 200).}}
\end{center}
\end{figure}

\begin{figure}[h!]
\begin{center}
\includegraphics[width=\textwidth]{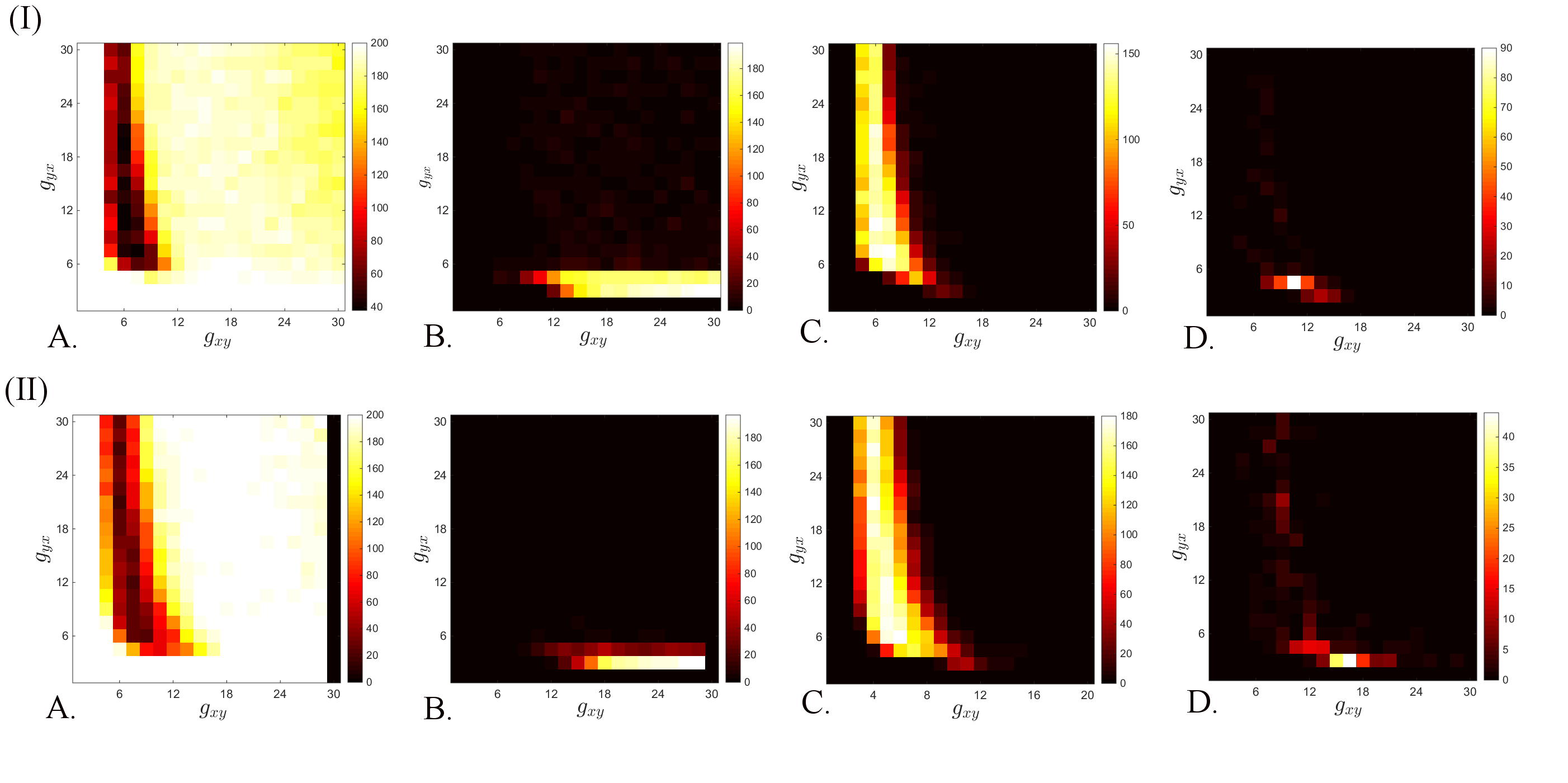}
\caption{\doublespacing \label{bifurcations_N4_different} \emph{\small  {\bf  Behavior frequency plots for $N=4$, for non-equal densities $M_{xy}$ and $M_{yx}$.} We show the fraction of all configurations in {\bf (i)} ${\cal D}^{6,10}$ and {\bf (ii)} ${\cal D}^{10,6}$ which exhibit: {\bf A.} one stable equilibrium; {\bf B.} one stable cycle; {\bf C.} multiple stable equilibria; {\bf D.} a coexisting stable cycle and equilibrium.  No other behaviors were found.  The simulations are based on a sample of $S=200$ configurations.}}
\end{center}
\end{figure}

 For example: in Figure~\ref{bifurcations_N2} we show, for $N=2$, the four nontrivial behavior frequency plots for ${\cal D}^{3,3}$.  These look as one would expect from ``overlapping'' the four dynamic classes in ${\cal D}^{3,3}$ (shown in Table~\ref{table_3_3}). Due to the similarities and differences between the Hopf and limit point bifurcation curves across configurations, the resulting frequency plots are a ``smeared'' version of the diagrams for individual classes. We conjecture that the profile of a frequency plot, as well as the degree of smearing (i.e., the width of the region with values transitioning between ${\cal P} = 0$ and ${\cal P} = 1$) depends on the pair $(M_{xy},M_{yx})$. Since there are such few different behaviors, one can still distinguish the contours of the individual bifurcation diagrams in the p-diagrams, which is no longer the case for higher $N$ (see Figures~\ref{bifurcations_N4_identical} and ~\ref{bifurcations_N4_different}). For larger $\lvert {\cal D}^{M_{xy},M_{yx}} \rvert$, there are more configurations, and more dynamic behaviors/classes. Since the number of configurations in  ${\cal D}^{M_{xy},M_{yx}}$ increases with $N$ extremely fast, when studying the same phenomena for larger values of $N$, it is more convenient to investigate the behavior distribution based on a sample probability. 

Figures~\ref{bifurcations_N4_identical} and ~\ref{bifurcations_N4_different} illustrate the behaviors we found numerically in a variety of systems with $N=4$. It is interesting to notice that we did not find aperiodic oscillations, or multiple stability. Each row shows, for one density type ($M_{xy},M_{yx})$, the frequency plots for the remaining four behaviors. Figures~\ref{bifurcations_N4_identical} illustrates how these loci change when the two densities $M_{xy}$ and $M_{yx}$ are identical, but increase from very low ($M_{xy}=M_{yx}=4$), to medium ($M_{xy}=M_{yx}=8$), to high ($M_{xy}=M_{yx}=12$). Figures~\ref{bifurcations_N4_different} illustrates two cases of uneven densities, one with $M_{xy}>M_{yx}$ and the other with $M_{yx}>M_{xy}$.

Broadly, one can notice that in some regions in the parameter plane the weights are a strong determinant of the potential dynamics, while in other regions only a very large jump in the $(g_{xy},g_{yx})$ parameter plane would significantly infuence the likely dynamics. The same applies to the sensitivity to weight changes: some regions are consistent between corresponding panels, showing that a switching from one density type to another would have almost no effect, while other regions are very sensitive to density and to configuration changes.

It is also interesting to notice that higer densities create sharper transitions, which is hardly surprising: if there are more edges, a small global change in the weights is more likely to have a substantial  effect on dynamics. It follows that, for lower densities, higher weights are required to place the system in an oscillatory (stable cycle) rather than quiet (stable equilibrium) regime. Moreover, the smoother spread of the plots for lower densities means that the dynamics is more susceptible to perturbations in configuration, even when the low densitites are fixed.\\

\noindent{\bf Remark 1.} Statistically speaking, the approach is appropriate when comparing behaviors within one single ${\cal D}^{M_{xy},M_{yx}}$, where each pair $(g_{xy},g_{yx})$ has the same number of corresponding configurations . When comparing behaviors between distributions ${\cal D}^{M_{xy},M_{yx}}$ for different values of $M_{xy}$ and $M_{yx}$, we tried to be more careful, and verified the validity of our sample-based method by computing the standard deviations over the chosen samples, to ensure that the results are not biased by using the same sample size for different size distributions.\\

\noindent {\bf Remark 2.} In the case of this simple system, the full-connectedness of the moduli maintains the moduli synchronized, so that, looking at the time evolution of one node, one can visualize with good approximation the temporal behavior of the whole module. This presents the advantage of behavior simple enough to be easily tractable even in higher dimensions. For example,  contrary to what one might have expected, our numerical searches did not find any parameter set for which the system exhibits aperiodic behavior. However, this is not a situation expected to occur biologically, and we use it only as a starting point. In the following section, we present an extension of this model which is a better candidate for biophisical and connectivity modeling in the brain, and which exhibits richer and more plausible behavior. We use the same techniques to investigate this extension.


\section{Coupled Wilson-Cowan pairs}
\label{coupled_WC_section}

We describe a more realistic scenario, where inhibition is implemented through a separate collection of nodes. This may be an appropriate representation of a brain network in which inhibition is performed via a hidden layer of neurons, different than the target cells that ultimately need to be inhibited. For example, the prefrontal cortex (PFC) projects excitatory fibers on the inter-neurons in ITC (an amygdala nucleus), which in turn inhibit the cells in the basal amygdala, the functional area considered to be responsible for emotion regulation. Hence the overall effect of the PFC on arousal reactions controlled by the amygdala produces  ``fear extinction'' (closing the negative feedback loop that regulates arousal).

To represent this situation, we consider the following model (see also Figure~\ref{coupled_WC_net}):

\begin{figure}[h!]
\begin{center}
\includegraphics[width=.96\textwidth]{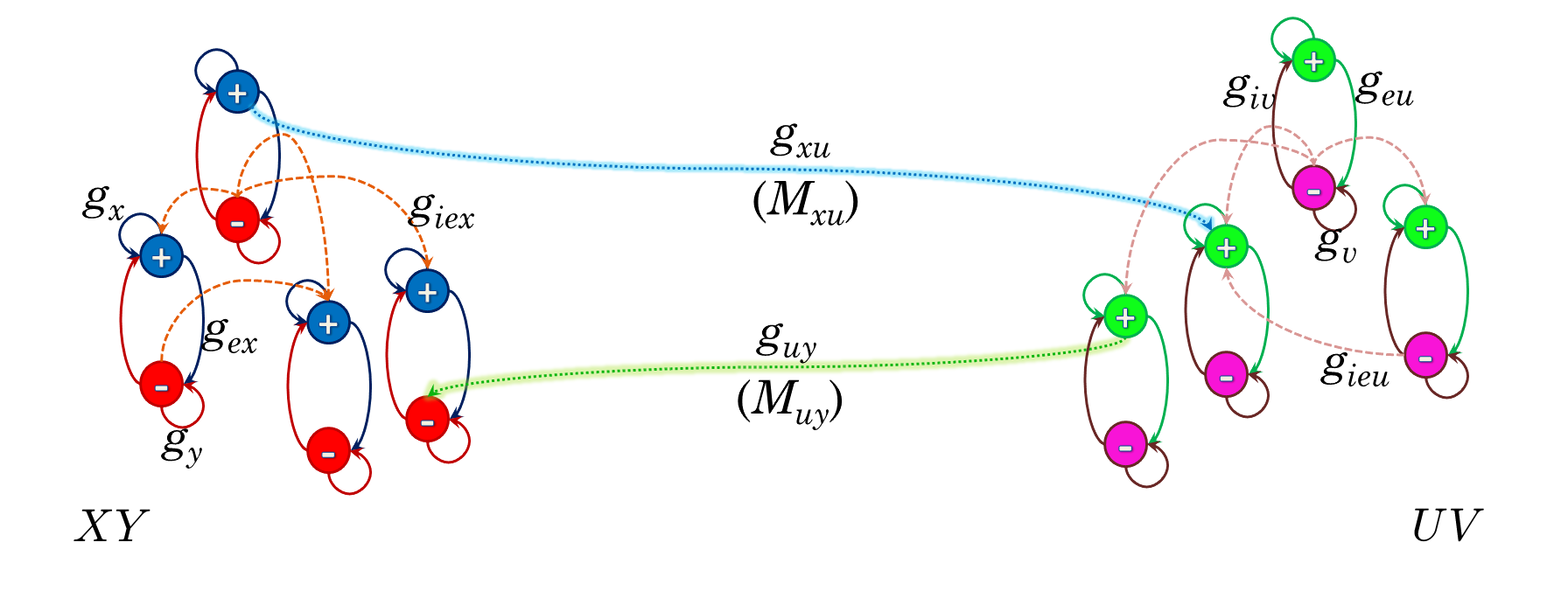}
\end{center}
\caption{\doublespacing \label{coupled_WC_net} \emph{\small {\bf A schematic representation of the coupled Wilson-Cowan system} for $N=4$ pairs of nodes in each module, $XY$ and respectively $UV$. Each (+)/(-) pair is coupled according to the original Wilson-Cowan model. In addition, each module has full (-) to (+) connectivity (i.e., each inhibitory unit is connected with all excitatory units within its module). A fraction $M_{xu}$ of the (+) units in module $XY$ are connected with (+) units in module $UV$, and a fraction $M_{uy}$ of (+) units in module $UV$ are connected with (-) units in module $XY$.}}
\end{figure}

\begin{eqnarray}
    \tau_e \frac{dx_k}{dt} &=& -x_k+(1-x_k) \cdot {\cal S}_{b_e,\theta_e} \left[ g_x x_k-g_{iy} y_k-\sum{g_{iex} y_p}+I \right] \nonumber \\
    \tau_i \frac{dy_k}{dt} &=& -y_k+(1-y_k) \cdot {\cal S}_{b_i,\theta_i} \left[ -g_y y_k+g_{ex} x_k+\sum{g_{uy} A_{kp}u_p} \right] \nonumber \\
    \tau_e \frac{du_k}{dt} &=& -u_k+(1-u_k) \cdot {\cal S}_{b_e,\theta_e} \left[ g_u u_k-g_{iv} v_k-\sum{g_{ieu}v_p}+\sum{g_{xu}B_{kp}x_p} \right] \nonumber \\
    \tau_i \frac{dv_k}{dt} &=& -v_k+(1-v_k) \cdot {\cal S}_{b_i,\theta_i} \left[ -g_v v_k+g_{eu} u_k \right]
\end{eqnarray}

\noindent  where ${\cal S}_{b,\theta}(\Sigma)=(1+\exp[-b(\Sigma-\theta)])^{-1}$, and we fixed the following Wilson-Cowan parameters: $b_e=1.3$, $b_i=2$, $\theta_e=4$, $\theta_i=3.7$, $I=1.5$. Connectivity parameters: $g_x=g_u=16$, $g_y=g_v=3$, $g_{ex}=g_{eu}=15$, $g_{iy}=g_{iv}=12$, $g_{iex}=g_{ieu}=5/N$, $g_{xu}=g_{uy}=10/N$.

While the modules retain a form of full (-) to (+) connectivity, the dynamics of this system is much more complex than that of the original model of simple coupled oscillators. Here, the network is spending most of its time in complex oscillatory regimes, in which the nodes are no longer synchronized within each module. We want to investigate whether this system exhibits aperiodic behavior, and what types of changes in the network configuration can throw the system from periodic oscillations into chaotic behavior.

\begin{figure}[h!]
\begin{center}
\includegraphics[width=\textwidth]{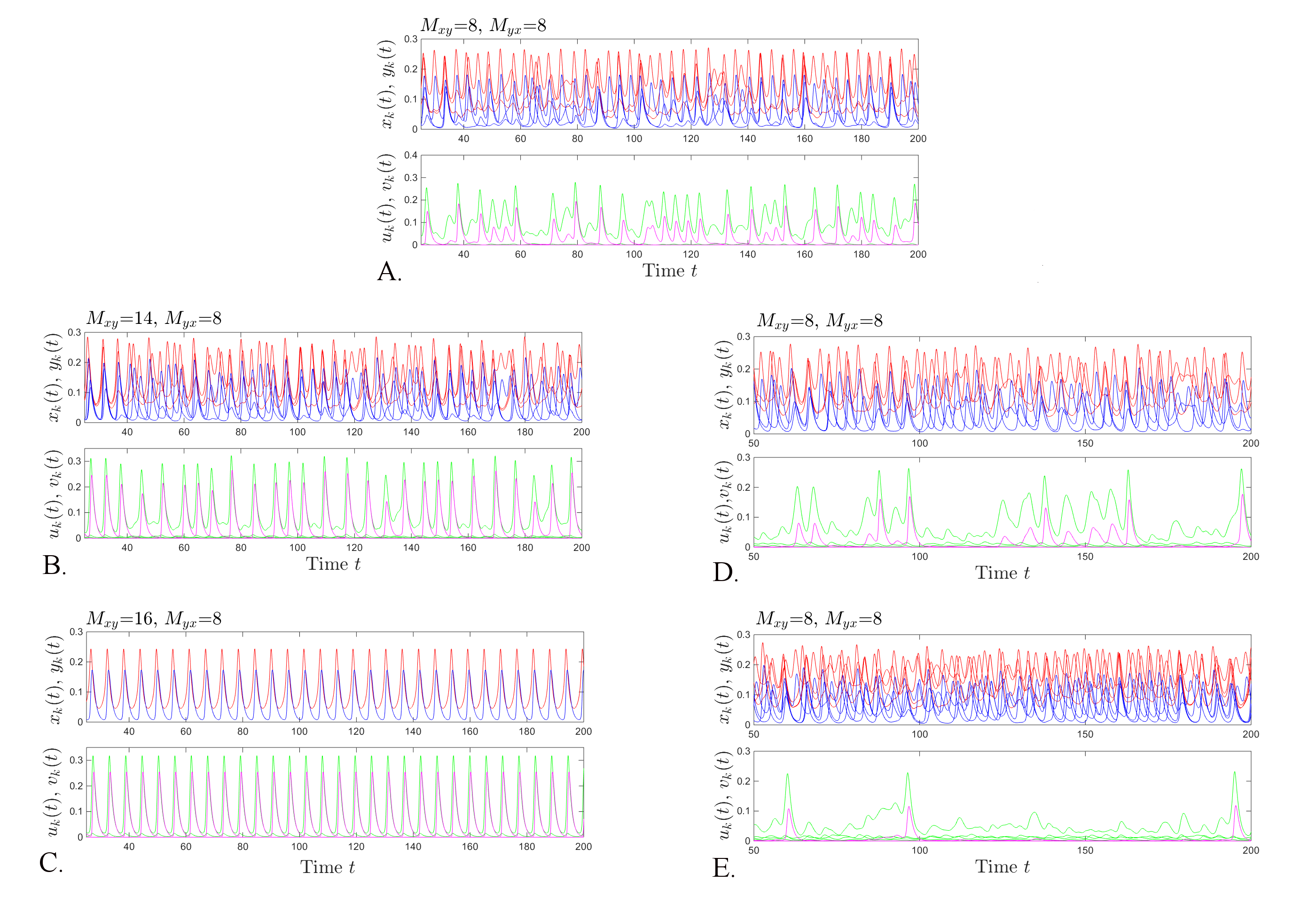}
\end{center}
\caption{\doublespacing \label{coupled_WC_traj}\emph{\small {\bf Transitions in dynamics when altering connectivity and configuration.} All panels show solutions for the Wilson-Cowan coupled pairs, for $N=4$ pairs of nodes in each module: the evolution of the nodes $x_k$ and $y_k$ is shown on top in blue and red, respectively, and the nodes $u_k$ and $v_k$ are shown on the bottom in green and purple. The simulations were performed for the parameters given in the text, and each for an arbitrary single configuration {\bf A.} of density type $(M_{xy},M_{yx})=(8,8)$; {\bf B.} of density type $(M_{xy},M_{yx})=(14,8)$; {\bf C.} of density type $(M_{xy},M_{yx})=(16,8)$; {\bf D.} of density type $(M_{xy},M_{yx})=(8,8)$, for a different adjacency configuration than that used in A; {\bf E.} of density type $(M_{xy},M_{yx})=(8,8)$, for a different configuration than those in A and D.}}
\end{figure}

Figure~\ref{coupled_WC_traj} shows, on the left, how the oscillatory regime can be affected by changes in density type. While for $(M_{xy},M_{yx})=(8,8)$ the nodes typically perform aperiodic oscillations, increasing $M_{xy}$ gradually introduces more structure (for $M_{xy}=14$) and renders them purely periodic (for $M_{xy}=16$). On the right, the figure illustrate how, for the same density pair, the oscillations can be tuned (singular ``spikes'' versus periodic ``bursts,'' versus sustained aperiodic oscillations) by altering only the configuration.

\begin{figure}[h!]
\begin{center}
\includegraphics[width=\textwidth]{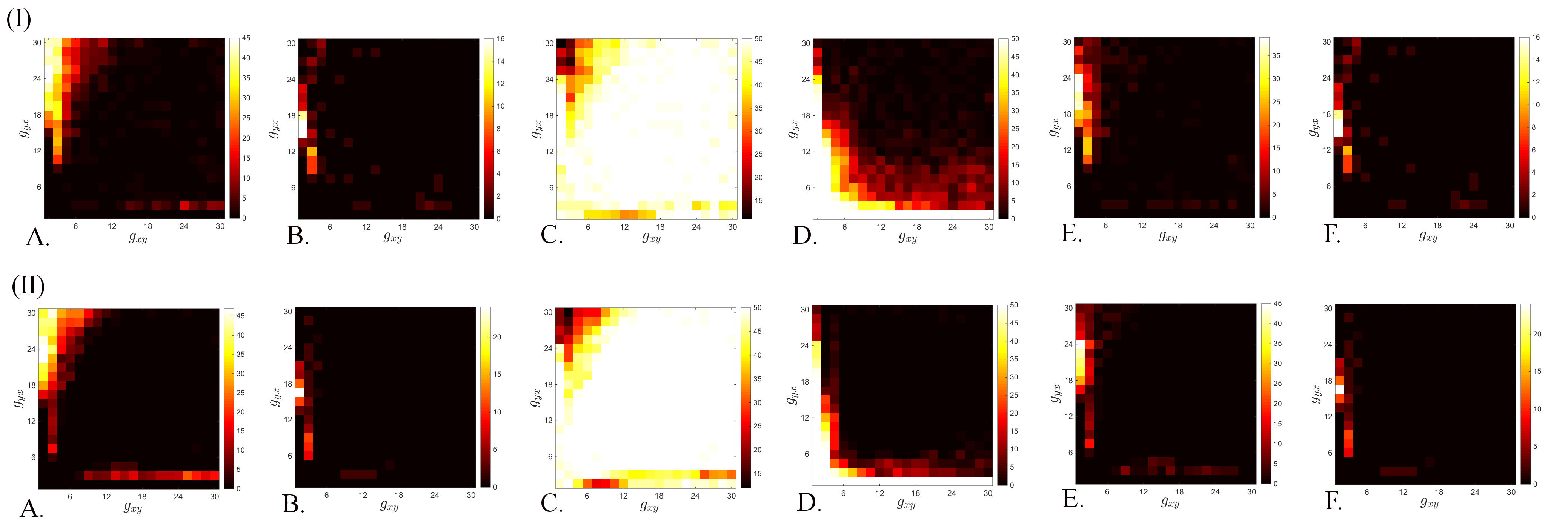}
\end{center}
\caption{\doublespacing \label{coupled_WC}\emph{\small {\bf Behavior frequency plots for the system of coupled Wilson-Cowan oscillators} for $N=4$, qnd density type ${\cal D}^{8,8}$ (top) and ${\cal D}^{14,8}$ (bottom). The illustrations are based on samples of size $S=50$. Each panel shows the number (out of the total of 50) of configurations leading to one of the following behaviors: {\bf A.} globally stable equilibrium; {\bf B.} multiple equilibria; {\bf C.} globally stable cycle; {\bf D.} aperiodic oscillations; {\bf E.} coexistance of a stable equilibrium with a stable cycle; {\bf F.} coexistence of multiple stable equilibria and cycles.}}
\end{figure}

As in the case for the simple coupled oscillators model, we aim to understand better the types of behaviors accessible to the system, and how changes in weights, densities or configuration may be used to swap between these behaviors. In Figure~\ref{coupled_WC}, we show the frequency plots in ${\cal D}^{8,8}$ and ${\cal D}^{14,8}$.

One imediately notices, in all both cases, the increased complexity of this system's dynamics compared with the simple model: all six behaviors appear in each of the density types, with large parameter loci allowing periodic and aperiodic oscillations. As in the previous system, however, richer behavior seems to correspond to lower densities (the higher the two densitites the more likely it is for the system to fall into simple periodic oscillations, as already suspected from Figure~\ref{coupled_WC_traj}).\\

\begin{figure}[h!]
\begin{center}
\includegraphics[width=.7\textwidth]{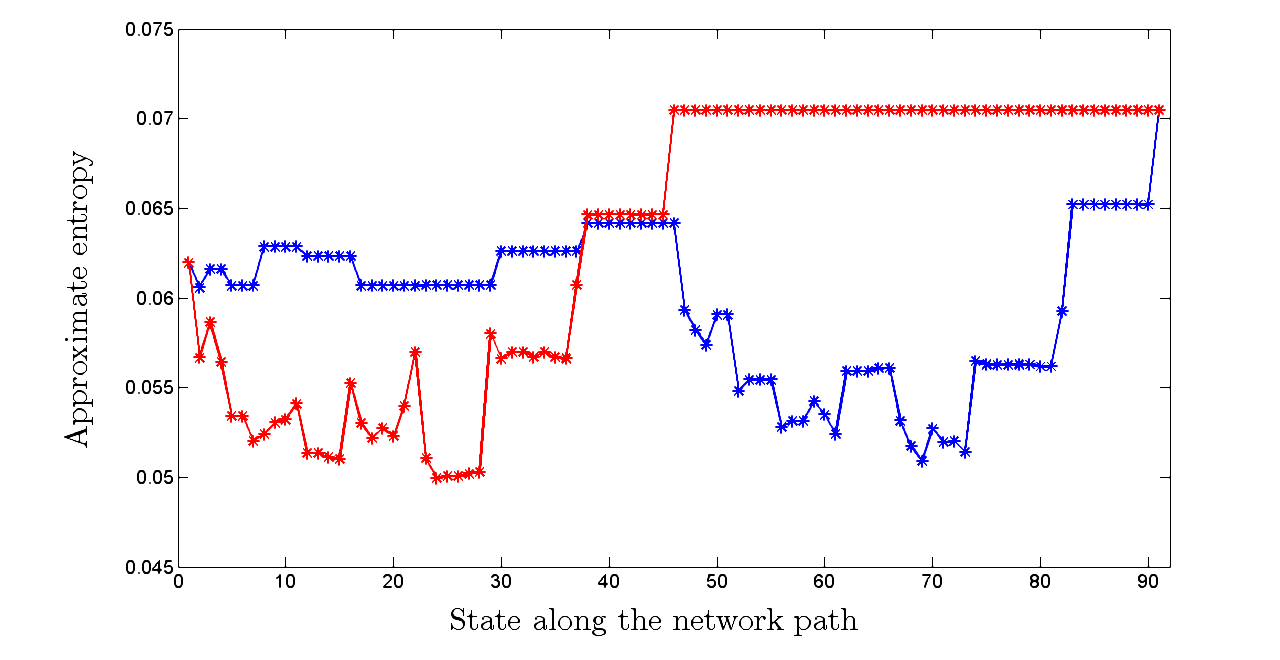}
\end{center}
\caption{\doublespacing \label{path_entropy} \emph{\small {\bf Evolution of the system's approximate entropy} from $h_0=0.0581$ to $h_1=0.0701$ along two distinct paths in the set of adjacency graph configurations. }}
\end{figure}

\noindent In Figure~\ref{path_entropy}, we illustrate one way of tracking changes in the system's dynamics when fixing the weights and density type and only changing the configuration by adding/deleting edges. The figure shows the evolution of the system's approximate entropy (estimated from the system's solutions according to an algorithm proposed by Pincus~\cite{pincus1991approximate}) along two network ``paths''  from one initial state (of relatively low entropy) to a final state (with higher entropy). More precisely, we considered an initial state in which only one unit in module $X$ is cross-connected to all units in module $Y$ (i.e., the block matrix $A$ has ones on the first row, and the block matrix $B$ has ones on the first column), and a final state in which the units are connected bijectively (both $A$ and $B$ are the identity). We constructed two paths in the adjacency graph from the initial to the final states, by defining each step to be a $0/1$ flip (a 1 swaps with a 0 at a neighboring position in the adjacency matrix, corresponding to an edge deletion and then addition in a proximal position). We want to suggest that there are many ways in which a system can evolve from low to high entropy through a sequence of slight edge perturbations (without the network even changing density type). We suspect that this is possible even if we additionally require the paths to be of monotonely increasing entropy. The states along these paths can be seen as states that the system will have to take provided it evolves along the respective path.

\section{Discussion}
\label{discussion}

\subsection{Strengthening versus restructuring}

In our paper, we focused on understanding a few aspects of how dynamic behavior in a network depends on its underlying adjacency matrix. To do this, we used an underlying graph with simple bimodular architecture, with each of the interconnected modules fully connected. We discovered that different temporal effects are to be expected when perturbing different aspects of the network connectivity. We compared the effects of globally increasing the \emph{weights} between the two interconnected modules versus increasing the \emph{number} of edges between the modules. While both actions lead to ``increasing connectivity'' between the two modules, they produced qualitatively different effects on dynamics. 

We noticed that, while certain regimes are robust to perturbations (local changes in weights or in adjacency don't produce qualitative effects on dynamics), other parameter regions tend to be very sensitive  to such changes. Furthermore, when in sensitive regimes, small \emph{local} perturbations in the network wiring (e.g., locally modifying the adjacency matrix by adding or deleting edges) may have dramatic effects on the system's dynamics, more substantial than those obtained by a \emph{global} change in the system's weights (recall that our weight parameters $g_{xy},g_{yx}$ affect all the connections from one module to another). 



Perhaps the most important question here is how the three hardware components (edge density, position and strength) act differently on the temporal behavior of the system, and how they work together to tune the network's dynamic complexity. This question is extremely important in the context of understanding a variety of real world networks.

\begin{figure}[h!]
\includegraphics[width=\textwidth]{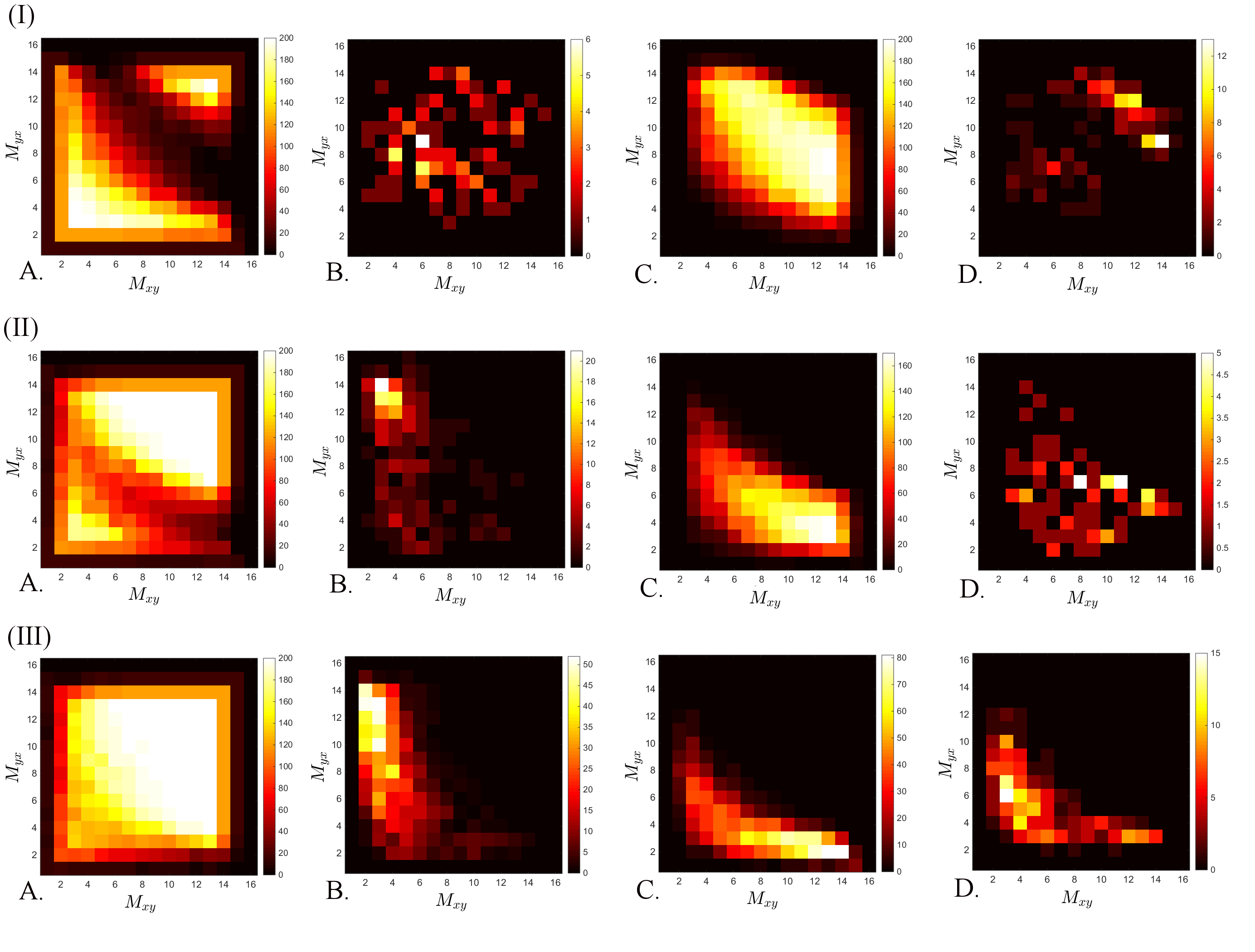}
\caption{\doublespacing \label{bifurcations_density} \emph{\small {\bf Behavior frequency plots in the parameter plane of densities $(M_{xy},M_{yx})$.} {\bf I.} for $g_{xy}=5$, $g_{yx}=10$; {\bf II.} for $g_{xy}=5$, $g_{yx}=60$; {\bf III.} for $g_{xy}=15$, $g_{yx}=15$.}}
\end{figure}

For example, one could think of what type of adjustments should be performed by the system in order to shift its dynamics most efficiently from a quiet to an oscillatory regime or vice-versa. If the state of the network is, to begin with, in a region sensitive to weight changes in the $(g_{xy},g_{yx})$ parameter plane, the system may perform the ``phase transition'' via a small change in the weights. Otherwise, if operating away from such regions, only a large, global change in the overall values of the weights can significantly increase the probability of the system to switch regimes. On the other hand, a small change in the graph structure could produce instead the desired dynamic change, pushing the system over into a more complex, or more stiff range of functioning. To help us illustrate this phenomenon, in Figure~\ref{bifurcations_density} we show the frequency plots for fixed weights $(g_{xy},g_{yx})$, with respect to the two densities $M_{xy}$ and $M_{yx}$, veiwed as system parameters. 

Consider for instance the situation in Figure~\ref{bifurcations_N4_identical}II(b), where it is clear that, for $M_{xy}=M_{yx}=8$, only a rather large change in weights would increase the potential for the quiet system  $g_{xy}=g_{yx}=15$ to oscillate. Figure~\ref{bifurcations_density}III (corresponding to $g_{xy}=g_{yx}=15$) shows, however, that the density point $(M_{xy},M_{yx})=(8,8)$ is in a sensitive region, where changes in the densities can strongly affect its likelihood to change behaviors, and changes in configuration can also be used to switch between available behaviors. For another example consider the point $g_{xy}=5$, $g_{yx}=10$, for which the system with $M_{xy}=10$ and $M_{yx}=6$ (shown in Figure~\ref{bifurcations_N4_different}II) performs oscillations with extremely high likelihood (for almost all configurations). Figure~\ref{bifurcations_density}I (for $g_{xy}=5$, $g_{yx}=10$) show that the point $M_{xy}=10$ and $M_{yx}=6$ is in a range where the dynamics is much more sensitive to changes in density and configuration.

Finally, recall that each density type generates a large number of distinct
configurations. The dynamics of the system may experience a whole collection of
dynamic modes (some qualitatively distinct, some equivalent) over the whole
distribution of possible configurations. While configuration-triggered changes in dynamics are important (and are in fact more likely in intermediate density type regimes), the dynamics seem, however, more robust to perturbations in configuration than to those involving a change in density type. This robustness, previously noticed in~\cite{radulescu2013network}, may be partly explained by the robustness of the adjacency spectrum when exploring all configurations of fixed density type~\cite{radulescu2014network}. However, in Section~\ref{hiDsection}, we have shown that adjacency spectrum classes are not in bijection with dynamics classes. Part of our current work is aimed towards understanding the theoretical bases of this robustness.


\subsection{Applications to learning and the brain connectome}

As discussed in a previous paper~\cite{radulescu2014network}, these choices of the mechanisms used to trigger changes in dynamics are extremely important for networked systems like the brain, in order to maintain their adequate function of performing complex simultaneous tasks. There are many different models describing the synaptic restructuring that occurs in a network of neurons during processes like learning, or memory formation, most likely involving a combination of weight changes of existing synapses, and creating/deleting connections. In terms of our model, this means that not only the edge weights, but also the edge distribution is likely to exhibit both short and long-term changes during learning. Knowledge of the geometry of the network  is therefore very important when determining which connectivity schemes are plausible to use for models of learning.

A lot of effort has been invested recently towards developing and using graph-theoretical network measures in conjunction with statistical methods, in order to identify the effects of abnormal connectivity patterns (measured as structural connectivity, for anatomical links; functional connectivity, for undirected statistical dependencies; and effective connectivity, for directed causal relationships among distributed responses~\cite{park2013structural}) on the efficiency of brain function.  By applying graph theoretical measures of segregation (e.g., clustering coefficient, motifs, modularity, rich clubs), integration (e.g., distance, path length, efficiency) and influence (e.g., node degree, centrality) these studies have been investigating the sensitivity of systems to removing/adding nodes or edges at different locations in the underlying network. 

Working with empirical data, such measures have been used to understand behavioral impairments in subjects with compromised connectivity due to existing lesions~\cite{corbetta2012functional}, or group differences between healthy controls and patients with mental illnesses associated with deficient feedback circuitry. In our previous work with fMRI data~\cite{radulescu2013network}, we ourselves used a simple graph-theoretical model as a formal framework to study how network density can affect the complexity of signal outputs,  measured by the log-log slope of their power spectra (power spectrum scale invariance, PSSI). Indeed, for sufficiently large networks, the log-log spectra were close to linear within certain frequency bands, and the PSSI slopes were found to vary as a function of both input type (excitatory, inhibitory) and input density (mean number of long-range connections), with comparatively insignificant dependence on the node-specific geometric distribution. 

Without attempting to understand the source of either dependence on density or robustness to specific configuration, we focused on the possible interpretations and applications. We suggested a testable framework for interpreting the empirical data in conjunction with the model, to deliver a connectivity-based hypothesis for the difference in functional regimes corresponding to different levels of anxiety. Individuals with average emotional reactivity had experimental  PSSI values in the pink noise range for amygdala and prefrontal regions, corresponding to well-regulated control systems, with well balanced excitatory and inhibitory projections. Individuals at the anxious end of the spectrum, showed experimentally white noise primarily for the amygdala, and were predicted by our model to have relatively weaker inhibitory inputs from the prefrontal cortex (producing weaker feedback). Individuals at the stress resilient end of the spectrum, showed white noise primarily for the prefrontal cortex, and were predicted by our model to have relatively stronger excitatory inputs from the amygdala (producing stronger feedback). This last simulation result may seem surprising, but in fact produces a reasonable hypothesis: enhanced projections from the amygdala to prefrontal cortex
effectively lower the threshold for inhibitory feedback, thereby suppressing all but the strongest stimuli. Broadly speaking, we saw as very promising the fact that such a simple and general setup may yet inform successfully our human imaging results in a circuit as important as the one regulating human emotion. That is because its simplicity allows us to study and understand (analytically or numerically) the sources that drive different aspects of the system's behavior (thus producing the different regimes of function); its generality opens such the model (with minor modifications) to possible applications other than emotion regulation.

The results in this paper (which used an identical network structure in its analysis) explain some of the more important (although perhaps counterintuitive) features observed computationally in R\v{a}dulescu et al~\cite{radulescu2013network}. Among these are the robustness of the coupled dynamics to certain changes in the network architecture and its vulnerability to others, as well as the differences between updating connection strengths versus perturbing connection density or geometry. 

In developing future iterations of this model with possible applications to learning mechanisms, it will also be important to explore how the learning process itself shapes the connectivity scheme, with possible emerging structures in which modularity is purposefully broken into hub-like subnetworks~\cite{siri2007effects}. Understanding the source and limits of a network's robustness and vulnerability to perturbations may be an instrument that could help us investigate in the future many aspects of brain circuitry: from determining which architectures favor convergence under particular learning algorithms, and which not, to classifying cognitive deficits and psychiatric illnesses.



\clearpage
\subsection*{Appendix: Adjacency versus dyamics classes for $N=2$}

\pagestyle{empty}

\begin{table}[h!]
\label{table_3_3}
\begin{minipage}{.5\textwidth}
\begin{center}
\begin{footnotesize}
\begin{tabular}{|c|c|}
\hline
& \\
$\;\left[ \begin{array}{cc|cc} &  & 1 & 1\\  &  & 1 & 0 \\ \cline{1-4} 1 & 1 &  & \\  1 & 0 &  & \end{array} \right] \; ({\cal A}_{iii})$ &
$\;\left[ \begin{array}{cc|cc} &  & 1 & 1\\  &  & 0 & 1 \\ \cline{1-4} 1 & 1 &  & \\  1 & 0 &  & \end{array} \right] \; ({\cal B}_{iv})$\\
& \\
\hline
& \\
$\;\left[ \begin{array}{cc|cc} & & 1 & 1\\  &  & 1 & 0 \\ \cline{1-4} 1 & 1 &  & \\  0 & 1 &  & \end{array} \right] \; ({\cal B}_{ii})$ &
$\;\left[ \begin{array}{cc|cc} & & 1 & 1\\  &  & 0 & 1 \\ \cline{1-4} 1 & 1 &  & \\  0 & 1 &  & \end{array} \right] \; ({\cal C}_{i})$\\
& \\
\hline
& \\
$\;\left[ \begin{array}{cc|cc} & & 1 & 1\\  &  & 1 & 0 \\ \cline{1-4} 1 & 0 &  & \\  1 & 1 &  & \end{array} \right] \; ({\cal B}_{iv})$ &
$\;\left[ \begin{array}{cc|cc} & & 1 & 1\\  &  & 0 & 1 \\ \cline{1-4} 1 & 0 &  & \\  1 & 1 &  & \end{array} \right] \; ({\cal A}_{iii})$\\
& \\
\hline
& \\
$\;\left[ \begin{array}{cc|cc} & & 1 & 1\\  &  & 1 & 0 \\ \cline{1-4} 0 & 1 &  & \\  1 & 1 &  & \end{array} \right] \; ({\cal C}_{i})$ &
$\;\left[ \begin{array}{cc|cc} & & 1 & 1\\  &  & 0 & 1 \\ \cline{1-4} 0 & 1 &  & \\  1 & 1 &  & \end{array} \right] \; ({\cal B}_{ii})$\\
& \\
\hline \hline
& \\
$\;\left[ \begin{array}{cc|cc} &  & 1 & 0\\  &  & 1 & 1 \\ \cline{1-4} 1 & 1 &  & \\  1 & 0 &  & \end{array} \right] \; ({\cal B}_{ii})$ &
$\;\left[ \begin{array}{cc|cc} &  & 0 & 1\\  &  & 1 & 1 \\ \cline{1-4} 1 & 1 &  & \\  1 & 0 &  & \end{array} \right] \; ({\cal C}_{i})$\\
& \\
\hline
& \\
$\;\left[ \begin{array}{cc|cc} & & 1 & 0\\  &  & 1 & 1 \\ \cline{1-4} 1 & 1 &  & \\  0 & 1 &  & \end{array} \right] \; ({\cal A}_{iii})$ &
$\;\left[ \begin{array}{cc|cc} & & 0 & 1\\  &  & 1 & 1 \\ \cline{1-4} 1 & 1 &  & \\  0 & 1 &  & \end{array} \right] \; ({\cal B}_{iv})$ \\
& \\
\hline
& \\
$\;\left[ \begin{array}{cc|cc} & & 1 & 0\\  &  & 1 & 1 \\ \cline{1-4} 1 & 0 &  & \\  1 & 1 &  & \end{array} \right] \; ({\cal C}_{i})$ &
$\;\left[ \begin{array}{cc|cc} & & 0 & 1\\  &  & 1 & 1 \\ \cline{1-4} 1 & 0 &  & \\  1 & 1 &  & \end{array} \right] \; ({\cal B}_{ii})$ \\
& \\
\hline
& \\
$\;\left[ \begin{array}{cc|cc} & & 1 & 0\\  &  & 1 & 1 \\ \cline{1-4} 0 & 1 &  & \\  1 & 1 &  & \end{array} \right] \; ({\cal B}_{iv})$ &
$\;\left[ \begin{array}{cc|cc} & & 0 & 1\\  &  & 1 & 1 \\ \cline{1-4} 0 & 1 &  & \\  1 & 1 &  & \end{array} \right] \; ({\cal A}_{iii})$\\
& \\
\hline
\end{tabular}
\end{footnotesize}
\end{center}
\end{minipage}
\quad
\begin{minipage}{.5\textwidth}
\begin{center}
\includegraphics[width=.55\textwidth]{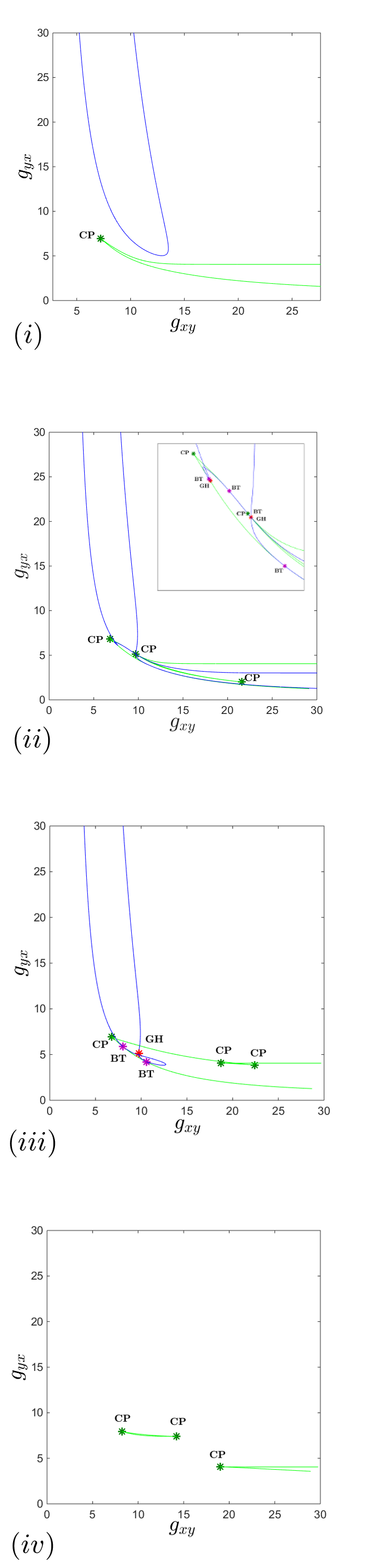}
\end{center}
\end{minipage}

\caption{\doublespacing \label{table_3_3} \emph{\small{\bf Adjacency and dynamics classes for N=2, density type ($M_{xy}$,$M_{yx}$)=(3,3).} Adjacency classes are designated by letters (${\cal A} - {\cal D}$) and dynamics classes by subscripts ($i - iv$).  The four possible parameter planes are shown on the right, with Hopf curves in blue, limit point curves in green and codimension two bifurcations marked with stars: cusp (green), Bautin (red) and Bogdanov-Takens (purple).}}
\end{table}

\begin{table}[h!]
\label{table_2_3}
\begin{footnotesize}
\begin{center}
\begin{tabular}{|c|c|c|c|}
\hline
& & & \\
$\;\left[ \begin{array}{cc|cc} &  & 1 & 1\\  &  & 0 & 0 \\ \cline{1-4} 1 & 1 &  & \\  1 & 0 &  & \end{array} \right] \; ({\cal A}_{v})$ &
$\;\left[ \begin{array}{cc|cc} & & 1 & 1\\  &  & 0 & 0 \\ \cline{1-4} 1 & 1 &  & \\  0 & 1 &  & \end{array} \right] \; ({\cal B}_{vi})$ &
$\;\left[ \begin{array}{cc|cc} & & 1 & 1\\  &  & 0 & 0 \\ \cline{1-4} 1 & 0 &  & \\  1 & 1 &  & \end{array} \right] \; ({\cal A}_{v})$ &
$\;\left[ \begin{array}{cc|cc} & & 1 & 1\\  &  & 0 & 0 \\ \cline{1-4} 0 & 1 &  & \\  1 & 1 &  & \end{array} \right] \; ({\cal B}_{vi})$\\
& & & \\
\hline
& & & \\
$\;\left[ \begin{array}{cc|cc} &  & 1 & 0\\  &  & 1 & 0 \\ \cline{1-4} 1 & 1 &  & \\  1 & 0 &  & \end{array} \right] \; ({\cal A}_{i})$ &
$\;\left[ \begin{array}{cc|cc} & & 1 & 0\\  &  & 1 & 0 \\ \cline{1-4} 1 & 1 &  & \\  0 & 1 &  & \end{array} \right] \; ({\cal A}_{i})$ &
$\;\left[ \begin{array}{cc|cc} & & 1 & 0\\  &  & 1 & 0 \\ \cline{1-4} 1 & 0 &  & \\  1 & 1 &  & \end{array} \right] \; ({\cal B}_{ii})$ &
$\;\left[ \begin{array}{cc|cc} & & 1 & 0\\  &  & 1 & 0 \\ \cline{1-4} 0 & 1 &  & \\  1 & 1 &  & \end{array} \right] \; ({\cal B}_{ii})$\\
& & & \\
\hline
& & & \\
$\;\left[ \begin{array}{cc|cc} &  & 1 & 0\\  &  & 0 & 1 \\ \cline{1-4} 1 & 1 &  & \\  1 & 0 &  & \end{array} \right] \; ({\cal C}_{iv})$ &
$\;\left[ \begin{array}{cc|cc} & & 1 & 0\\  &  & 0 & 1 \\ \cline{1-4} 1 & 1 &  & \\  0 & 1 &  & \end{array} \right] \; ({\cal D}_{iii})$ &
$\;\left[ \begin{array}{cc|cc} & & 1 & 0\\  &  & 0 & 1 \\ \cline{1-4} 1 & 0 &  & \\  1 & 1 &  & \end{array} \right] \; ({\cal D}_{iii})$ &
$\;\left[ \begin{array}{cc|cc} & & 1 & 0\\  &  & 0 & 1 \\ \cline{1-4} 0 & 1 &  & \\  1 & 1 &  & \end{array} \right] \; ({\cal C}_{iv})$\\
& & & \\
\hline
& & & \\
$\;\left[ \begin{array}{cc|cc} &  & 0 & 1\\  &  & 1 & 0 \\ \cline{1-4} 1 & 1 &  & \\  1 & 0 &  & \end{array} \right] \; ({\cal D}_{iii})$ &
$\;\left[ \begin{array}{cc|cc} & & 0 & 1\\  &  & 1 & 0 \\ \cline{1-4} 1 & 1 &  & \\  0 & 1 &  & \end{array} \right] \; ({\cal C}_{iv})$ &
$\;\left[ \begin{array}{cc|cc} & & 0 & 1\\  &  & 1 & 0 \\ \cline{1-4} 1 & 0 &  & \\  1 & 1 &  & \end{array} \right] \; ({\cal C}_{iv})$ &
$\;\left[ \begin{array}{cc|cc} & & 0 & 1\\  &  & 1 & 0 \\ \cline{1-4} 0 & 1 &  & \\  1 & 1 &  & \end{array} \right] \; ({\cal D}_{iii})$\\
& & & \\
\hline
& & & \\
$\;\left[ \begin{array}{cc|cc} &  & 0 & 1\\  &  & 0 & 1 \\ \cline{1-4} 1 & 1 &  & \\  1 & 0 &  & \end{array} \right] \; ({\cal B}_{ii})$ &
$\;\left[ \begin{array}{cc|cc} & & 0 & 1\\  &  & 0 & 1 \\ \cline{1-4} 1 & 1 &  & \\  0 & 1 &  & \end{array} \right] \; ({\cal B}_{ii})$ &
$\;\left[ \begin{array}{cc|cc} & & 0 & 1\\  &  & 0 & 1 \\ \cline{1-4} 1 & 0 &  & \\  1 & 1 &  & \end{array} \right] \; ({\cal A}_{i})$ &
$\;\left[ \begin{array}{cc|cc} & & 0 & 1\\  &  & 0 & 1 \\ \cline{1-4} 0 & 1 &  & \\  1 & 1 &  & \end{array} \right] \; ({\cal A}_{i})$\\
& & & \\
\hline
& & & \\
$\;\left[ \begin{array}{cc|cc} &  & 0 & 0\\  &  & 1 & 1 \\ \cline{1-4} 1 & 1 &  & \\  1 & 0 &  & \end{array} \right] \; ({\cal B}_{vi})$ &
$\;\left[ \begin{array}{cc|cc} & & 0 & 0\\  &  & 1 & 1 \\ \cline{1-4} 1 & 1 &  & \\  0 & 1 &  & \end{array} \right] \; ({\cal A}_{v})$ &
$\;\left[ \begin{array}{cc|cc} & & 0 & 0\\  &  & 1 & 1 \\ \cline{1-4} 1 & 0 &  & \\  1 & 1 &  & \end{array} \right] \; ({\cal B}_{vi})$ &
$\;\left[ \begin{array}{cc|cc} & & 0 & 0\\  &  & 1 & 1 \\ \cline{1-4} 0 & 1 &  & \\  1 & 1 &  & \end{array} \right] \; ({\cal A}_{v})$ \\
& & & \\
\hline
\end{tabular}

\vspace{3mm}
\includegraphics[width=.85\textwidth]{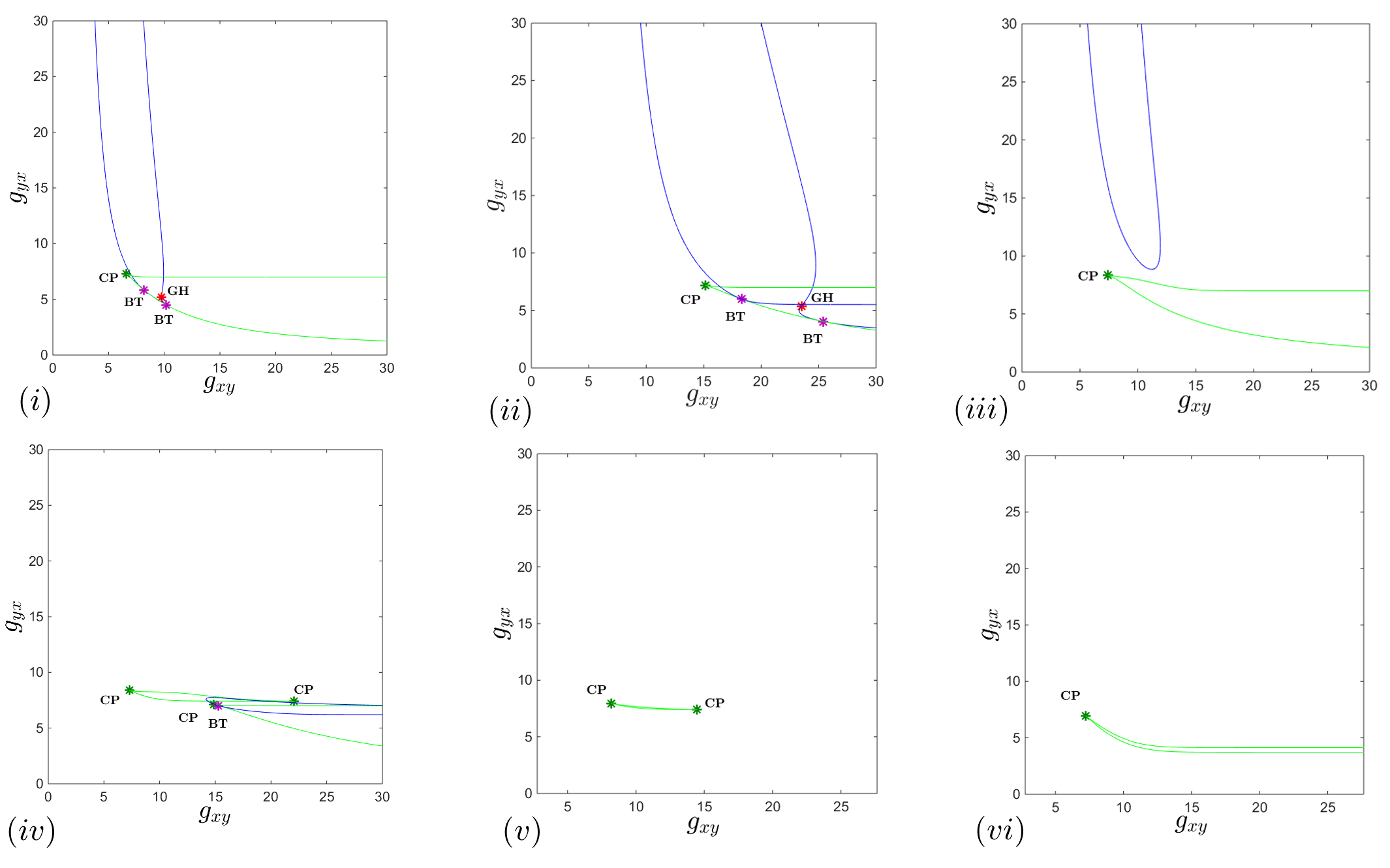}
\end{center}
\end{footnotesize}

\caption{\doublespacing \label{table_2_3} \emph{\small{\bf Adjacency and dynamics classes for N=2, density type ($M_{xy}$,$M_{yx}$)=(2,3).}}}
\end{table}


\end{document}